\documentclass[letterpaper]{article} 
\usepackage[draft]{aaai2026}  
\usepackage{times}  
\usepackage{helvet}  
\usepackage{courier}  
\usepackage[hyphens]{url}  
\usepackage{graphicx} 
\def\UrlFont{\rm}  
\usepackage{natbib}  
\usepackage{caption} 
\usepackage{algorithm}
\usepackage{booktabs}
\usepackage{graphicx}
\usepackage{latexsym}
\usepackage{amsmath}
\usepackage{multirow}
\usepackage{array}
\usepackage{comment}
\usepackage{algorithm}
\usepackage{algpseudocode}
\usepackage[tikz]{mdframed}
\usepackage{anyfontsize}
\usepackage{sidecap}
\usepackage{rotating}
\newcolumntype{C}[1]{>{\centering\arraybackslash}p{#1}}
\newcolumntype{L}[1]{>{\raggedright\arraybackslash}p{#1}}

\usepackage{newfloat}
\usepackage{listings}
\DeclareCaptionStyle{ruled}{labelfont=normalfont,labelsep=colon,strut=off} 
\floatstyle{ruled}
\newfloat{listing}{tb}{lst}{}
\floatname{listing}{Listing}
\title{HierSearch: A Hierarchical Enterprise Deep Search Framework Integrating Local and Web Searches}
\author{
    Jiejun Tan\textsuperscript{\rm 12}\thanks{This work was done when Jiejun Tan was doing an internship at Baichuan.},
    Zhicheng Dou\textsuperscript{\rm 1}\thanks{Corresponding author.}, 
    Yan Yu\textsuperscript{\rm 2},
    Jiehan Cheng\textsuperscript{\rm 12},
    Qiang Ju\textsuperscript{\rm 2},
    Jian Xie\textsuperscript{\rm 2},
    Ji-Rong Wen\textsuperscript{\rm 1} \\
}
\affiliations{
    \textsuperscript{\rm 1}Gaoling School of Artificial Intelligence, Renmin University of China \\
    \textsuperscript{\rm 2}Baichuan Intelligent Technology \\

    \{zstanjj, dou, jrwen\}@ruc.edu.cn
}

\usepackage{bibentry}

\begin{document}

\maketitle

\begin{abstract}
Recently, large reasoning models have demonstrated strong mathematical and coding abilities, and deep search leverages their reasoning capabilities in challenging information retrieval tasks. Existing deep search works are generally limited to a single knowledge source, either local or the Web. However, enterprises often require private deep search systems that can leverage search tools over both local and the Web corpus. Simply training an agent equipped with multiple search tools using flat reinforcement learning (RL) is a straightforward idea, but it has problems such as low training data efficiency and poor mastery of complex tools.
To address the above issue, we propose a hierarchical agentic deep search framework, HierSearch, trained with hierarchical RL. At the low level, a local deep search agent and a Web deep search agent are trained to retrieve evidence from their corresponding domains. At the high level, a planner agent coordinates low-level agents and provides the final answer. Moreover, to prevent direct answer copying and error propagation, we design a knowledge refiner that filters out hallucinations and irrelevant evidence returned by low-level agents.
Experiments show that HierSearch achieves better performance compared to flat RL, and outperforms various deep search and multi-source retrieval-augmented generation baselines in six benchmarks across general, finance, and medical domains.~\footnote{Code and datasets are available at \url{https://github.com/plageon/HierSearch}}
\end{abstract}


\section{Introduction}
Recently, large reasoning models~(LRMs) such as DeepSeek-R1~\cite{deepseek-r1} and OpenAI's O-series~\cite{o3-o4} models have shown impressive capabilities in mathematics and coding. However, LRMs are troubled by higher hallucination rates~\cite{o3-hallucination,r1-hallucination,lrm-hall-detection} and restricted by limited internal knowledge in knowledge-intensive tasks. Thus, studies have combined LRMs with retrieval-augmented generation (RAG) to enable models to obtain external knowledge assistance, which is referred to as deep search~\cite{search-o1,webthinker}.

Existing deep search works often equip LRMs with a local corpus search tool~\cite{recall,deepseek-r1,r1-searcher} or a Web search tool~\cite{search-o1,webthinker,deep-researcher}. However, a common scenario for most enterprises is that their private deep search system interacts with both local knowledge sources and Web knowledge sources~\cite{multi-source-pruning}. To be specific, enterprises often possess private domain-specific documents. Existing methods for building private RAG systems usually involve processing them into a text chunk corpus and constructing knowledge graphs~\cite{ms-graph-rag,lightrag,medrag}. Web knowledge sources generally include search engines and web pages. Generally speaking, local knowledge sources are more professional and targeted. Meanwhile, Web knowledge sources are more comprehensive and timely~\cite{prefrag,astute-rag}. This deep search scenario with multiple knowledge sources poses challenges to existing methods: \textit{Deep search agents need to selectively use different knowledge sources based on user questions and the characteristics of knowledge sources, and cross-supplement missing knowledge.}

A straightforward solution for the above challenge is equipping the deep search agent with all search tools for all knowledge sources and conducting flat reinforcement learning~(RL). However, the flat RL solution is not suitable for the following reasons: (1) Numerous search tools result in a large action space during RL, leading to low training efficiency and instability. 
(2) Search tools within the same knowledge source have stronger synergy~(e.g., browsing a Web page via a URL retrieved by a search engine or retrieving text chunks mentioning an entity from the knowledge graph), while that between tools across different knowledge sources is weaker. However, flat RL fails to effectively utilize this characteristic.
(3) Moreover, preliminary experiments show that during flat RL, rewards encourage the agent to search more frequently in easily retrievable knowledge sources, while less frequently in hard ones~(Web search is more difficult in our setting due to a wider search scope and more noise). Thus, the training efficiency of flat RL for the difficult knowledge source is poor due to limited exploration of the corresponding tools. 

To address the above issues, we propose a hierarchical agentic deep search paradigm, HierSearch, which comprises a local deep search agent, a Web deep search agent, and a planner agent. Two deep search agents interact directly with search tools within their knowledge sources and retrieve evidence for the planner agent. Specifically, the local deep search agent has access to the local text chunk corpus and the local knowledge graph. The Web deep search agent has access to the Web search engine and online web pages. Meanwhile, the planner agent drafts search plans, coordinates search agents, analyzes evidence provided by search agents, and provides the final answer.

Accordingly, we leverage a hierarchical reinforcement learning~(HRL)~\cite{hrl-survey} algorithm to train this hierarchical agentic framework. Also, we use Group Relative Policy Optimization~(GRPO)~\cite{grpo} and rule-based rewards. HRL overcomes the challenges above, mainly manifested in:
(1) \textbf{In the first stage, we train low-level agents, the local deep search agent and the Web deep search agent separately.} They master search tools within the same domain well, because the number of tools is limited and the tools are closely related. 
(2) \textbf{In the second stage, we train the high-level planner agent, equipped with both deep search agents.} Well-trained deep search agents mask the complex interaction process with search tools, and greatly lower the difficulty of knowledge acquisition. The planner agent can learn search planning and knowledge integration across multiple knowledge sources faster and better.

In the planner agent's training stage, we find that directly providing the complete trajectories of deep search agents would introduce irrelevant search results and the agents' hallucinatory reasoning contents. To address this, we design a reasoning-aware knowledge refiner. This refiner first selects the evidence that contributes to each round of reasoning by the deep search agent. Second, it selects the evidence helpful to the agent's conclusion from an overall perspective.

We conduct extensive experiments on six benchmarks from the general domain, the medical domain, and the financial domain. The results show that HierSearch outperforms baselines and the flat RL solution across all benchmarks.

In summary, our contributions are threefold: (1) We explore the deep search framework in multi-knowledge-source scenarios and propose a hierarchical agentic paradigm and train with HRL; (2) We notice drawbacks of the naive information transmission among deep search agents and developed a knowledge refiner suitable for multi-knowledge-source scenarios; (3) Our proposed approach for reliable and effective deep search across multiple knowledge sources outperforms existing baselines the flat-RL solution in various domains.

\section{Related Works}

\subsubsection{Deep Search} Traditional RAG combines large language models~(LLMs) with information retrieval to provide external knowledge and mitigate hallucination~\cite{trust-rag-survey,rag-survey,flashrag}. Traditional RAG methods generally follow a fixed retrieve-then-generate pipeline~\cite{dparag_dgt,htmlrag,flashrag}, while several works explore flexible agenitic pipelines~\cite{self-rag,react}. Compared to traditional RAG, deep search combines LRM with search tools~\cite{rag-reasoning-servey,synergizing-rag}. Equipped with stronger reasoning abilities, deep search pushes iterative RAG further, and analyzes deeper for users' questions~\cite{webthinker}, which can ``\textit{search, read and reason until best answer found}''~\cite{jinaai-deepsearch}.
Several organizations have developed their representative products, such as Google~\cite{gemini-deepsearch}, OpenAI~\cite{openai-deepsearch}, and Jina~\cite{jinaai-deepsearch}.
Meanwhile, several researchers builds deep search on open-source LRMs~\cite{deepseek-r1,qwq32b,qwen3-tech-report}, like RAG-Star~\cite{rag-star}, Search-o1~\cite{search-o1}, and WebThinker~\cite{webthinker}. These works often have issues of excessive reasoning and inaccurate searching in search tasks, but they have the advantage of greater flexibility in choosing models and search tools~\cite{rearag,agentic-reasoning,manusearch}.
To make reasoning models perform better in deep search tasks, another branch of works train LLMs to conduct deep search tasks following the RL paradigm introduced by DeepSeek-R1~\cite{deepseek-r1}, like DeepResearcher~\cite{deep-researcher}, R1-Searcher~\cite{r1-searcher}, and ReCall~\cite{recall}. The aforementioned deep search works are all limited to a single knowledge source, and at most two search tools~\cite{search-r1,zero-search}. However, enterprise private deep search often needs to access local and Web knowledge sources as well as multiple search tools. Existing methods cannot supplement knowledge and handle knowledge conflicts across multiple knowledge sources. Moreover, they produce a lot of unnecessary search tool calls, especially expensive Web search tool calls. In contrast, HierSearch uses multiple deep search agents to tackle different knowledge sources, and a planner agent that selectively calls agents of different knowledge sources as needed and integrates knowledge from these sources.

\subsubsection{Multi-Knowledge Source RAG}
In traditional RAG research, some works have identified the challenges RAG faces in multi-knowledge-source scenarios and proposed solutions.
PruningRAG~\cite{multi-source-pruning} uses multi-granularity pruning strategies to integrate information from documents of different sources and mitigate the impact of misleading information.
PrefRAG~\cite{prefrag} introduces preference-driven adaptive retrieval to handle multi-retrieval source data, and calls web retrieval as a supplement when local retrieval does not satisfy knowledge requirements. 
HM-RAG~\cite{hm-rag} applies multi-source agents to conduct retrieval in parallel, and uses consistency voting to integrate multi-source answers.
These works are still static RAG paradigms that need to follow a predefined pipeline. They use prompting or DPO methods to enable agents to learn multi-source RAG tasks. In contrast, we apply the GRPO RL algorithm to develop the agent's deep thinking and searching capabilities. Agents with deep thinking capabilities are not limited to a specific search path; instead, they can independently plan when to call search tools, when to interact with other agents, and when to terminate.

\begin{figure*}
    \centering
    \includegraphics[width=\linewidth]{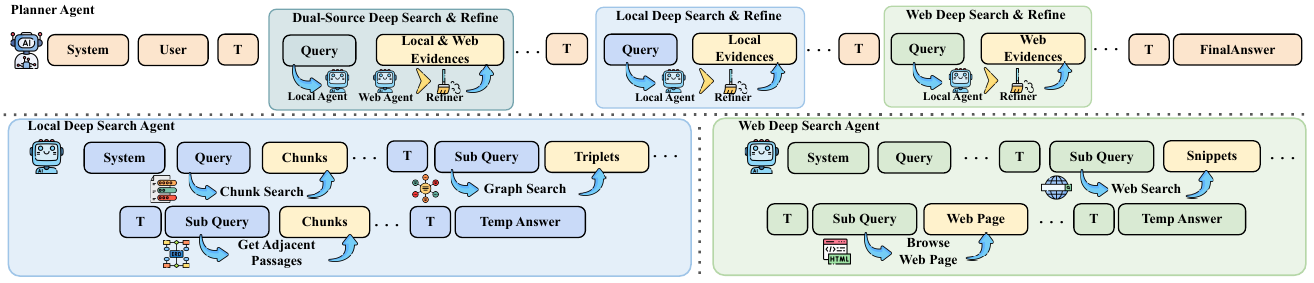}
    \caption{Illustration of the hierarchical agentic framework for HierSearch. We show exemplary trajectories of all low-level and high-level agents.}
    \label{fig:pipeline}
\end{figure*}

\subsubsection{Hierarchical RL}
HRL decomposes complex tasks into simpler subtasks and uses a high-level policy to select subtasks and a low-level policy to perform specific actions~\cite{feudal-hrl,feudal-rl,maxq}. HRL is effective and data-efficient when used in tasks with multiple turns, long horizons, and delayed rewards~\cite{hrl-survey,hrl-survey-1}.
HRL has performed well in robot control and game AI~\cite{hiro,deep-hrl,hrl-intrin}. Recent works have also applied HRL to agents built on LLMs~\cite{archer,epo}. To the best of our knowledge, this work is the first to use HRL in the deep search field. Multi-knowledge-source RAG task is broken down into two levels: in-knowledge-source deep search and cross-knowledge-source planning.

\section{Methodology}
We present HierSearch, a hierarchical agentic framework designed for enterprise-wide deep search across multiple knowledge sources. The framework comprises two levels: 1) low-level agents, including local and Web deep search agents, and 2) a high-level planner agent.

\subsection{Problem Formulation}

In a deep search task, the agent takes a user's question $x$, iteratively performs thinking processes or search tool calls, and finally outputs an answer $\hat{y}$. The optimization goal is to make the final answer as correct and helpful as possible. In the enterprise scenario, a deep search needs to access multiple knowledge sources before providing an answer. Given knowledge sources including a local text chunk corpus $C$, a local knowledge graph $G$, a Web search engine $E$, and accessible Web pages on the Internet $P$, the deep search framework is meant to maximize the probability of the golden answer $y$.

\subsection{Hierarchical Agentic Deep Search}
A straightforward idea for the multi-knowledge-source deep search task is equipping an agent with all search tools and conducting flat RL. However, our preliminary experiment shows that the flat RL displays drawbacks such as poor mastery of difficult Web search tools and low training data efficiency. Thus, we propose a hierarchical agentic deep search framework, HierSearch. As shown in Figure~\ref{fig:pipeline}, HierSearch consists of a local deep search agent, a Web deep search agent, and a planner agent. We will discuss all three agents in the following sections in detail, including their accessible tools and their roles.

\subsubsection{Preliminary: Tool-Augmented Reasoning}
We follow a commonly used synergized tool-augmented reasoning paradigm of current deep search methods~\cite{rag-reasoning-servey}. Our deep search agents and the planner agent roll out similarly. We use the following wrapping tags to distinguish different part in the trajectory: (1) The thinking processes are wrapped in \texttt{$<$think$>$}...\texttt{$<$/think$>$}; (2) Tool calls are wrapped in \texttt{$<$tool\_name$>$}...\texttt{$<$/tool\_name$>$}~(The tool name varies). (3) Returned contents tool functions are wrapped in \texttt{$<$result$>$}...\texttt{$<$/result$>$}. (4) The answer is wrapped in \texttt{$<$answer$>$}...\texttt{$<$/answer$>$}. All tools accessible are demonstrated in the system prompt. The generation process pauses when the ending tags of tool calls are generated, and restarts until the tool call result is appended to the end of the sequence. The whole generation process ends when \texttt{$<$/answer$>$} is generated or the number of tool call rounds reaches an upper limit.

\subsubsection{Local Deep Search Agent}

The local deep search agent has access to two local knowledge sources: the text chunk corpus and the knowledge graph. The local agent accesses the text chunk corpus through \texttt{$<$chunk\_search$>$} to retrieve chunks related to the input query. The local agent accesses knowledge graph by two tools: (1)~\texttt{$<$graph\_search$>$} retrieves triples (consisting of a subject, a predicate, and an object) related to the input query by calculating the similarity of semantic embeddings; (2)~\texttt{$<$get\_adjacent\_passages$>$} returns relevant text chunks mentioning the input entity in the knowledge graph. The linking edges between graph entities and relevant chunks are identified and saved during the knowledge graph construction process.

\begin{figure*}
    \centering
    \includegraphics[width=\linewidth]{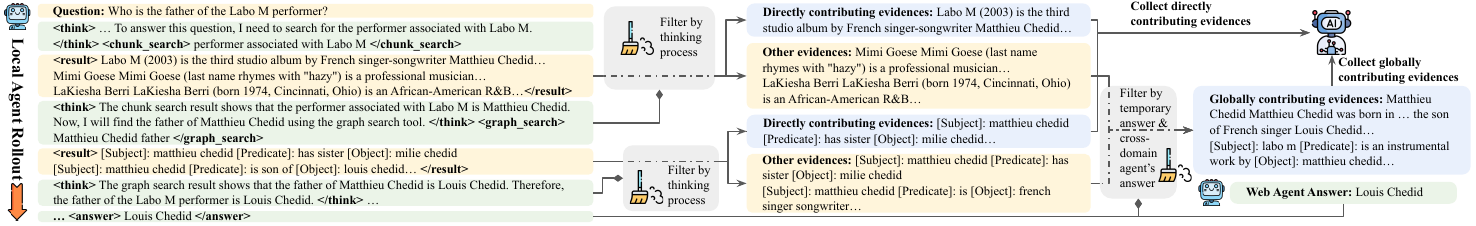}
    \caption{Illustration of the knowledge refining process from the local agent trajectory. The first step filters directly contributing evidence according to the subsequent thinking process of each round. The second step filters globally contributing evidence according to the local agent's answer and the web agent's answer (if available).}
    \label{fig:refiner}
\end{figure*}

\subsubsection{Web Deep Search Agent}

The Web deep search agent accesses Web knowledge through two tools: (1)~\texttt{$<$web\_search$>$} calls a search engine API to retrieve web links and corresponding titles and snippets related to the input query; (2)~\texttt{$<$browse\_url$>$} takes both a web link and a query as inputs. We chunk the original HTML pages and only return query-relevant pieces, because the original HTML pages are generally lengthy and hard to read.

\subsubsection{Multi-Knowledge Source planner agent}
Both the local deep search agent and the Web deep search agent are low-level agents that are manipulated by a high-level planner agent.
The planner agent drafts search plans, integrates returned evidence from low-level agents, and provides the final answer. Low-level agents are packaged as tools for high-level agents to call, which includes the following: (1)~\texttt{$<$local\_search\_agent$>$} calls the local deep search agent; (2)~\texttt{$<$web\_search\_agent$>$} calls the Web deep search agent; (3)~\texttt{$<$all\_search\_agent$>$} calls both low-level agents simultaneously.

\subsubsection{Hierarchical RL for Multi-Source Deep Search}

Considering the hierarchical framework and taking inspiration from HRL works, we employ HRL for HierSearch. That is, we first train two low-level search agents, and then the high-level planner agent. To be specific, we randomly sample the training set from MuSiQue~\cite{musique}, OmniEval~\cite{omnieval}, and BioASQ~\cite{bioasq}. We mix these samples as the training data for agents.

We follow the GRPO algorithm introduced by DeepSeek-R1~\cite{deepseek-r1}, and we use rule-based rewards, which are designed as follows. Agent trajectories with incorrect formats are punished with a zero reward. If the format is correct, we calculate the F1 score between the predicted answer $\hat{y}$ and the golden answer $y$. If the F1 score is larger than zero, we take the F1 score as the reward. If the rollout has a correct format but a zero F1 score, we encourage the agent to explore more tools. We calculate the proportion of the types of tools used during the rollout to the total types of tools accessible to the agent, and multiply it by a coefficient of 0.1 to serve as the reward. To sum up, the reward function can be formulated as:
\begin{equation}
    R =
    \begin{cases}
    0, & \text{if the format is incorrect,} \\
    0.1 \times t / T, & \text{if F1 = 0 and format is correct,} \\
    \text{F1}(\hat{y},y), & \text{if F1 } > \text{0 and format is correct.}
    \end{cases}
\end{equation}
, where $t$ is the number of tools used in the trajectory and $T$ is the number of all tools accessible.

\begin{table*}[t]
\small
\centering
\setlength{\tabcolsep}{1.8mm}{
\begin{tabular}{lcccccccccccc|cc}
\toprule
\multirow{2}{*}{\textbf{Method}} & \multicolumn{2}{c}{\textbf{MuSiQue}} & \multicolumn{2}{c}{\textbf{OmniEval}} & \multicolumn{2}{c}{\textbf{BioASQ}} & \multicolumn{2}{c}{\textbf{NQ}} & \multicolumn{2}{c}{\textbf{HotpotQA}} & \multicolumn{2}{c|}{\textbf{PubmedQA}} & \multicolumn{2}{c}{\textbf{\# Searches}} \\
\cmidrule(lr){2-3}\cmidrule(lr){4-5}\cmidrule(lr){6-7}\cmidrule(lr){8-9}\cmidrule(lr){10-11}\cmidrule(lr){12-13}\cmidrule(lr){14-15}
& EM & F1 & EM & F1 & EM & F1 & EM & F1 & EM & F1 & EM & F1 & Local & Web\\
\midrule
\multicolumn{15}{l}{\textbf{Local Search}} \\
DeepSeek-R1   & 26.00 & 36.45 &  0.80 & 29.50 &  6.18 & 24.10 & 28.50 & 44.88 & 29.75 & 45.36 & 41.00 & 51.35 & 2.00 &   -  \\
HippoRAG      & 30.25 & 43.36 &  0.00 & 29.27 &  9.71 & 36.87 & 43.25 & 59.71 & 35.25 & 52.23 & 68.50 & 70.95 & 2.00 &   -  \\
R1-Searcher   & 44.50 & 55.86 &  2.93 & 9.85 & 34.12 & 50.87 & 44.50 & 56.92 & 48.25 & 63.88 & 64.00 & 64.12  & 1.68 &   -  \\
ReCall        & 42.75 & 53.82 &  8.53 & 23.01 & 24.71 & 43.30 & 47.50 & 61.09 & \underline{49.50} & \underline{63.99} & 28.00 & 34.64 & 2.55 &   -  \\
\midrule
\multicolumn{15}{l}{\textbf{Web Search}} \\
DeepSeek-R1   & 22.50 & 32.60 &  0.53 & 24.23 &  5.29 & 20.25 & 26.75 & 39.89 & 26.50 & 40.31 & 15.25 & 30.22 &   -  & 1.00 \\
DeepResearcher& 30.00 & 39.44 &  2.40 & 17.95 & 28.82 & 46.80 & 41.50 & 54.99 & 39.50 & 52.95 & 56.25 & 56.79 &   -  & 2.84 \\
Search-o1     & 28.50 & 39.03 &  3.20 & 15.37 & 30.59 & 47.24 & 36.00 & 48.79 & 42.00 & 53.80 & 64.00 & 67.19 &   -  & 1.72 \\
WebThinker    & 30.75 & 42.15 &  1.33 & 15.90 & 33.24 & 49.82 & 36.75 & 50.52 & 43.50 & 58.68 & 65.00 & 66.07 &   -  & 2.55 \\
\midrule
\multicolumn{15}{l}{\textbf{Parallel Search}} \\
DeepSeek-R1   & 26.50 & 37.47 &  1.07 & 28.31 &  4.41 & 22.34 & 23.75 & 39.51 & 28.50 & 44.37 & 40.25 & 50.13 & 2.00 & 1.00 \\
HippoRAG      & 33.25 & 46.39 &  0.00 & 29.69 & 10.29 & 37.29 & 43.00 & 59.88 & 39.75 & 57.70 & 70.25 & 70.93 & 2.00 & 1.00 \\
HM-RAG        & 26.25 & 37.59 &  7.73 & 35.93 & 13.53 & 39.01 & 43.75 & 59.76 & 44.00 & 59.50 & \underline{71.25} & \underline{71.29} & 5.27 & 2.64 \\
R1-Searcher   & \underline{46.50} & \underline{57.19} &  2.67 &  9.22 & 33.82 & 50.54 & 44.75 & 56.97 & 47.75 & 62.93 & 66.25 & 66.52 & 3.36 & 1.68 \\
ReCall        & 43.00 & 52.69 &  9.33 & 22.02 & 26.18 & 42.45 & \underline{48.25} & \underline{61.13} & 47.00 & 62.12 & 31.75 & 39.34 & 4.62 & 2.31\\
DeepResearcher& 33.75 & 44.94 &  6.40 & 24.96 & 32.94 & 52.44 & 46.25 & 59.76 & 45.75 & 60.23 & 64.75 & 65.50 & 4.20 & 2.10 \\
Search-o1     & 36.25 & 47.53 &  5.60 & 18.82 & 32.06 & 50.18 & 39.25 & 53.36 & 44.00 & 59.13 & 65.50 & 68.93 & 3.10 & 1.55 \\
WebThinker    & 33.00 & 44.53 &  5.60 & 19.87 & 33.82 & 50.54 & 40.25 & 53.39 & 46.75 & 61.04 & 67.75 & 69.04 & 4.38 & 2.19 \\

\midrule
\multicolumn{15}{l}{\textbf{Selective Search}} \\
CRAG          & 26.50 & 36.89 &  1.07 & 28.50 &  5.88 & 23.76 & 25.00 & 42.25 & 30.00 & 45.43 & 41.50 & 51.76 & 2.00 & 0.61 \\
PrefRAG       & 33.75 & 47.47 &  \underline{9.60} & 40.19 & 11.18 & 38.47 & 40.00 & 57.01 & 43.50 & 61.56 & 60.25 & 65.29 & 2.18 & 0.04 \\
HierSearch$_\text{w/o HRL}$ & 46.00 & 56.34 &  7.73 & \underline{39.49} & \underline{39.41} & \underline{62.42} & 47.75 & 59.65 & 42.00 & 57.99 & 67.50 & 69.31 & 4.82 & 1.02 \\
HierSearch       & \textbf{53.00} & \textbf{62.83}  & \textbf{10.67}  & \textbf{46.37} & \textbf{49.94}  & \textbf{66.99} & \textbf{57.00}  &\textbf{68.00}  & \textbf{53.25} & \textbf{67.40}& \textbf{71.75} & \textbf{72.81} & 5.74 & 1.06 \\
\bottomrule
\end{tabular}
}
\caption{Main Results of HierSearch. The best and second best of each model are in \textbf{bold} and \underline{underlined}.}
\label{tab:main-results}
\end{table*}

\subsection{Reasoning-Aware Knowledge Refiner}

This hierarchical framework requires information exchange between low-level deep search agents and the high-level planner agent. A straightforward idea is that low-level agents return the whole trajectory containing collected evidence (search results from search tools), thinking processes, and conclusions (temporary answers in answer tags). However, analytical experiments show that inputting all those information indiscriminately will be harmful for the planner, which mainly shows in: (1) Thinking processes and conclusions from low-level agents induce the planner agent directly copy them instead of thinking by itself; (2) Irrelevant evidence makes the contents low-level agents' returned lengthy and hard to read and confuses the planner agent; (3) The hallucinations generated by low-level agents lead to an error propagation to the planner agent.

Therefore, we design a knowledge refiner that filters key evidence contributing to the low-level agents' thinking processes and conclusions, as shown in Figure~\ref{fig:refiner}.
The refiner filters evidence helpful for the thinking process in two steps. In the first refining step, we select evidence directly contributing to the next thinking process. Given a trajectory sequence $S$, which contains an input question $x$, and $K$ rounds where thinking and tool calls alternate, and ends with a last thinking process $t_{K+1}$ followed by a conclusion $\hat{c}$. The round $k$ contains a thinking process $t_k$, a query $q_k$, and $N$ returned evidence $\{e_{N(k-1)+1} \cdots e_{Nk}\}$. The trajectory sequence is like:
\begin{equation}
    S = \{x,t_1, \cdots, t_k, q_k, e_{N(k-1)+1} \cdots e_{Nk}, \cdots, t_{K+1}, \hat{c}\}
\end{equation}
The contribution score for each evidence in round $k$ is given by its contribution to the next thinking process:
\begin{equation}
    \text{Score}(e_i) = P(e_i|t_{k+1}),\text{  }N(k-1)+1\leq i\leq Nk
\end{equation}
The contribution score is calculated by the embedding similarity score $P$ between the evidence and the subsequent thinking process. In the first step, in each think \& search round, top $\alpha$\% evidence is selected.

In the second refining step, we distinguish evidence not selected in the first step but contributing globally to the agent's conclusion. As preparation, unselected evidence after the first step is gathered as candidates. 
If the planner agent calls only one low-level agent, we consider only that low-level agent's conclusion $\hat{c}$. If the planner agent calls both low-level agents, we concatenate $\hat{c}$ with the other agent's conclusion $\hat{c}^{'}$ as $\{\hat{c},\hat{c}^{'}\}$, and consider them as a whole.
The global contribution score for the conclusion is given by:
\begin{equation}
    \text{Score}(e_i) = 
    \begin{cases}
    
    P(e_i|\{\hat{c},\hat{c}^{'}\}), & \hat{c}^{'} \text{exists,} \\
    P(e_i|\hat{c}),&  \text{otherwise.}
    \end{cases}
\end{equation}
In the second step, the top $\beta$\% of the remaining evidence is selected. Both $\alpha$ and $\beta$ are hyperparameters of the refiner. Finally, evidence selected from the two steps is merged and tagged with its knowledge source.
The planner receives a list of refined evidence collected by agents (e.g., ``$<$result$>$ Local Knowledge Graph: [Subject] matthieu chedid ... Search Engine: Labo M (2003) is the third studio album ... $<$/result$>$'').

\section{Experiments}

\subsection{Benchmarks}
We select three general-domain benchmarks, including: (1)~MuSiQue~\cite{musique}: A synthetic multi-hop QA dataset; (2)~Natural Questions~(NQ)~\cite{nq}: Real search engine questions collected by Google; (3)~HotpotQA~\cite{hotpot-qa}: A multi-hop QA dataset based on Wikipedia. We select one financial-domain benchmark, OmniEval~\cite{omnieval}, a Chinese large-scale RAG benchmark targeting the financial domain with human annotations. We select two medical-domain benchmarks: (1)~BioASQ~\cite{bioasq}: An annually updated biomedicine challenge with QA tasks; (2)~PubMedQA~\cite{pubmedqa}: A human-annotated QA dataset based on research papers on PubMed. All benchmarks in finance and biomedicine include numerous queries that can only be answered using local knowledge. We randomly sample 373 samples for OmniEval, 340 samples for BioASQ, and 400 samples for other benchmarks from their corresponding test set (if available). We calculate Exact Match (EM) and F1 score as evaluation metrics for all benchmarks. Also, we count the average local search and Web search times (Web page browsing not included) required to process a query for each method.

\begin{table*}[t]
\small
\centering
\setlength{\tabcolsep}{1.2mm}{
\begin{tabular}{lcccccc}
\toprule
\textbf{Method} & \textbf{MuSiQue} & \textbf{OmniEval} & \textbf{BioASQ} & \textbf{NQ} & \textbf{HotpotQA} & \textbf{PubmedQA} \\
\midrule
HierSearch       & \textbf{53.00} & \textbf{10.67}  & \textbf{49.94}  & \textbf{57.00}  & \textbf{53.25} & \textbf{71.75}     \\

w/o Local Agent       & 29.75 (23.25\%$\downarrow$) &  3.20  (7.47\%$\downarrow$) & 35.00 (14.94\%$\downarrow$) & 36.00 (21.00\%$\downarrow$) & 33.25 (20.00\%$\downarrow$) & 65.00  (6.75\%$\downarrow$) \\
w/o Web Agent         & 47.50  (5.50\%$\downarrow$) &  9.87  (0.80\%$\downarrow$) & 46.18  (3.76\%$\downarrow$) & 55.50  (1.50\%$\downarrow$) & 51.50  (1.75\%$\downarrow$) & 69.50  (2.25\%$\downarrow$) \\
w/o Refiner           & 50.75 (2.25\%$\downarrow$) &  9.60  (1.07\%$\downarrow$) & 48.82  (1.12\%$\downarrow$) & 56.25  (0.75\%$\downarrow$) & 48.50  (4.75\%$\downarrow$) & 68.50  (3.25\%$\downarrow$) \\

\bottomrule
\end{tabular}
}
\caption{Ablation Study.}
\label{tab:ablation-study}
\end{table*}

\subsection{Baselines}
To demonstrate the effectiveness of our method, we select the following baselines:

$\bullet$ \textbf{Local Search}. (1) HippoRAG\cite{hipporag}: The graph RAG backbone method, with GPT-4o-mini~\cite{4o-mini} as the base model. (2) DeepSeek-R1~\cite{deepseek-r1}: A powerful reasoning model augmented by single-time chunk search and graph search; (3) R1-Searcher~\cite{r1-searcher} and (4) Recall~\cite{recall}: Both are deep search agents trained from scratch on QA datasets in local retrieval environments.

$\bullet$ \textbf{Web Search}. (1)~A powerful reasoning model augmented by single-time Web search; (2)~Search-o1~\cite{search-o1}: A deep search method that incorporates Web search into reasoning in a single inference chain; (3)~WebThinker~\cite{webthinker}: A deep search method that involves a deep web explorer in a main reasoning chain;
(4)~DeepResearcher~\cite{deep-researcher}: A deep search agent trained from scratch in real-world web environments.

$\bullet$ \textbf{Parallel Search}. To align the knowledge sources and make a fair comparison, we reproduce the above baselines in a parallel search setting, where the same query is sent to both local and Web search tools in parallel. Also, we reproduce HM-RAG~\cite{hm-rag}, which conducts parallel RAG based on text search, graph search, and Web search, and merges three answers with a majority vote.

$\bullet$ \textbf{Selective Search}. The agent autonomously decide which knowledge source to use or both, including: (1) PrefRAG~\cite{prefrag}: A multi-turn RAG pipeline that decides wether to involve Web search basing on local retrieval results; (2) CRAG~\cite{crag}: A plug-in discriminator that decides using Web search, local search or both basing on local retrieval results; (3) HierSearch$_\text{w/o HRL}$: A deep search agent equipped with all search tools and trained by flat RL.

\subsection{Implementation Details}
For local search, we prepare local knowledge bases separately for general, medical, and financial domains. For the general domain, we sample passages from the Wikipedia dump, and for the medical domain, we sample passages from the PubMed dump. 
The sampling passages consist of directly related passages for questions and hard negatives retrieved by BM25~\cite{bm25}. This corpus sampling process for Wikipedia and PubMed is necessary because their original sizes are too large for constructing a graph upon them. For the financial domain, we use the original retrieval corpus of OmniEval.
The knowledge graph is constructed upon the text chunk corpus. We follow HippoRAG~\cite{hipporag} and employ GPT-4o-mini~\cite{4o-mini} and BGE-M3~\cite{bge-m3} in graph construction. BGE-M3 is also the embedding model for all local search tools.
As for the Web search, \texttt{$<$web\_search$>$} uses Bing Search API for English queries and the Quark Search API for Chinese queries. \texttt{$<$browse\_url$>$} accesses real-time Web pages and extract relevant evidence.
For training settings, we collect training samples from Musique, OmniEval, and BioASQ. We train the local deep search agent, Web deep search agent, and the planner agent for 300 steps with a batch size of 64 and \texttt{Qwen2.5-7B-Instruct}~\cite{qwen2.5-tech-report} as the backbone. More implementation details are in the appendix.

\subsection{Main Results}
Main experimental results are shown in Table~\ref{tab:main-results}. ``\# Searches'' is the average local or Web search tools (Web page browsing excluded) called to search for a user's question. Experiments demonstrate that HierSearch outperforms baselines without many additional search tool calls. Additionally, we make the following observations: (1) Baseline methods generally perform better if they have access to more knowledge sources. Local search has a larger augmentation than Web search because they are more professional and targeted.
(2) Compared to methods using parallel search to access multiple knowledge sources, our method exhibits stronger deep search capabilities in multi-knowledge-source environments. Also, parallel search methods generate more Web search tool calls, which are slow and expensive.
(3) Compared to multi-knowledge-source RAG baselines using selective search, our method is not constrained by a fixed workflow in knowledge source selection and integration, and makes deeper search and thinking. As for the comparison to flat RL~(HierSearch$_\text{w/o HRL}$), we make a detailed analysis in the section below ``\textit{Effectiveness of Hierarchical RL}''.
(4) NQ, HotPotQA, and PubmedQA are not included in our training data, so the performances on them demonstrate our method's generalization ability in out-of-scope scenarios.

\subsection{Further Analysis}
\subsubsection{Ablation Study}
We conduct an ablation study on key modules of our method, as shown in Table~\ref{tab:ablation-study}. We make the following observations:
(1) We ablate the local deep search agent. In practice, we return an empty result when the planner calls the local deep search agent. Due to the lack of local information, the ultimate performance decreases. 
(2) Similar to the local deep search agent, when the Web deep search agent is ablated, the ultimate performance decreases due to a lack of Web knowledge.
(3) We ablate the knowledge refiner. In practice, we directly provide the planner with the complete content of agent trajectories. Since the trajectories contain irrelevant search results and hallucinations from low-level agents, the overall performance is affected.

\subsubsection{Effectiveness of Hierarchical RL}

\begin{figure}[t]
    \centering
    \includegraphics[width=\linewidth]{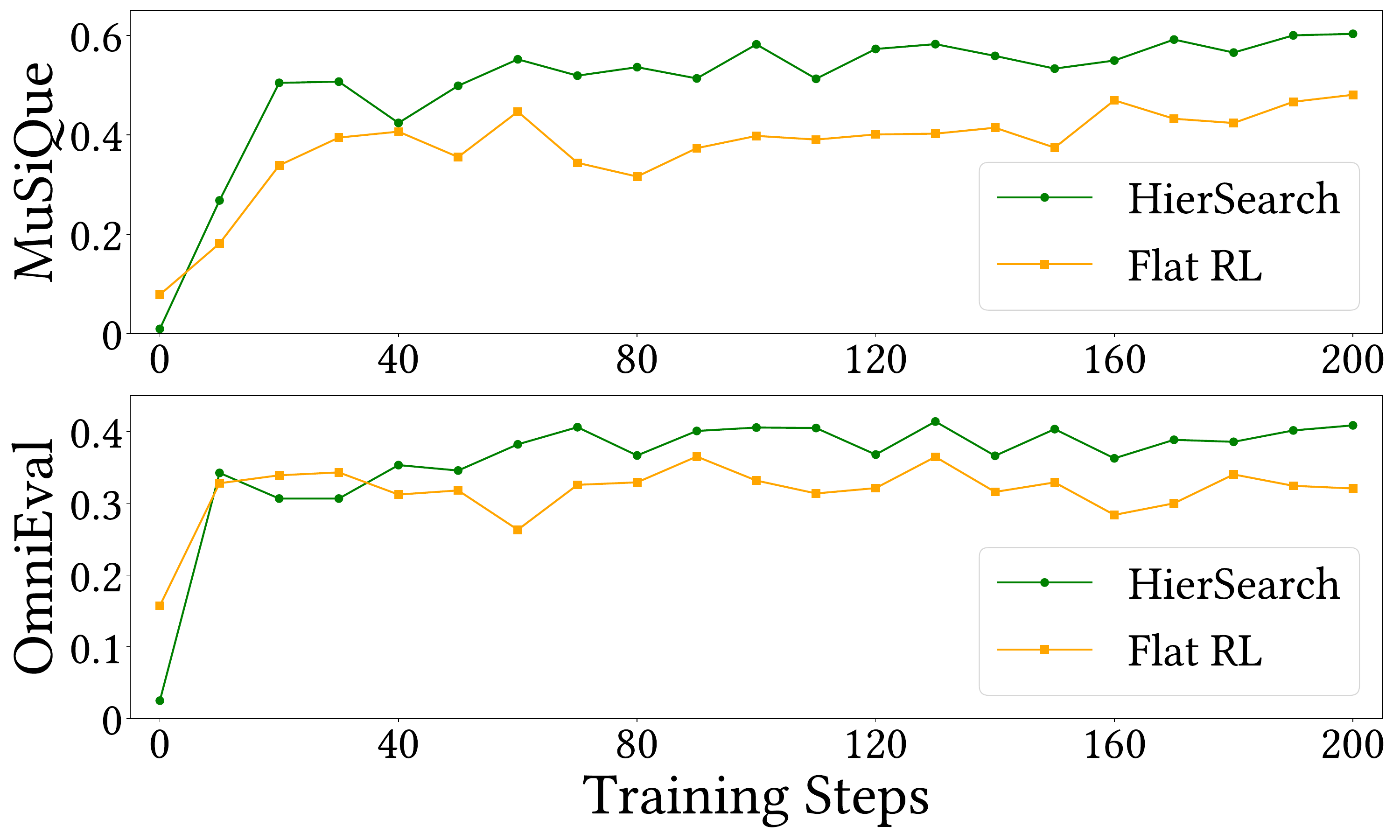}
    \caption{Rewards on Validation Sets during Training.}
    \label{fig:training-steps}
\end{figure}

To demonstrate that HRL has an edge over flat RL under the multi-knowledge-sources environment, we conduct a training comparison experiment. We start from an identical backbone model, \texttt{Qwen2.5-7B-Instruct}, and train with identical training samples. The training batch size is set to 64, and the total number of training steps is 300. We evaluate the checkpoint every 10 steps during training with a validation set sampled from MuSiQue and OmniEval. The results for the first 200 steps are shown in Figure~\ref{fig:training-steps}~(due to space limitations). The green curve represents HierSearch using HRL, while the orange curve represents the flat-RL-trained agent. Comparing the reward curves, we can see:
(1)~At the initial stage, both methods' performances grow rapidly due to learning the tool-calling format, and the performance of HierSearch grows faster than flat RL. 
(2)~Both methods' performance enters a plateau on MuSiQue after 20 steps, which is the same on OmniEval after 10 steps. During this period (steps 20 to 300), both methods are improving their deep search abilities slowly with fluctuations, and HierSearch consistently performs better than flat RL. 

Additionally, through case analysis (more details in the Appendix), we find that:  
(1) The strong performance of HierSearch benefits from the low-level agents' stronger deep search ability compared to original search tools, as well as the refiner's ability to refine key evidence. 
(2) Flat RL faces multiple search tools and a larger action space, resulting in low sample utilization efficiency. Further analysis shows that at step 290, the Web search tools only account for 18.5\% of the total search tool calls, leading to low optimization efficiency.

\subsubsection{Analysis of Multi-Knowledge Source Searching}

To further reveal the detailed reason that our method performs better in multi-knowledge-source environments, we analyze search success rates and reasoning success rates, and identify both of them according to different knowledge sources. To be specific, if the gold answer is contained in returned local search results, it is regarded as a successful local search. This also applies to Web search, and ``both'' means both local and Web search are successful. The search success rate is calculated by dividing the number of search successful samples by the total number of samples. Meanwhile, under the premise of a successful local search, if the planner agent gives a correct final answer (EM $=$ 1), it is regarded as successful reasoning. The reasoning success rate is calculated by dividing the number of reasoning successful samples by the number of search successful samples, which are the same for the Web and both. 

Results are shown in Table~\ref{tab:success-rate}, and we make the following observations: (1) Local search is easier than Web search, while Web search supplements some knowledge. (2) Specialized deep search agents have a higher search and reasoning success rate than agents built on general reasoning models. (3) Compared to deep search baselines, our method is better at searching as well as reasoning. (4) The flat RL solution (HarmoS$_\text{w/o HRL}$) outperforms all baselines in local search success rate and is close to our hierarchical method. However, its performance in web search success rate is unsatisfactory. This confirms our observation in preliminary experiments: the flat RL solution insufficiently explores and optimizes web search tools.

\subsubsection{Efficiency Analysis}

Since we employ a hierarchical framework consisting of three agents, which may raise efficiency concerns, we make a comprehensive computational efficiency analysis, as shown in Table~\ref{tab:efficiency}. We count the number of local search tool calls (\# LS), Web search tool calls (\# WS), Web page browsing tool calls (\# WB), reasoning tokens (\# Tokens), and the overall latency.  
For parallel search baselines, we call graph search, chunk search, and Web search tools in parallel whenever the agent provides a query. The first two are local search tools, so their local search count is exactly twice that of Web search.  
For latency calculation, we estimate it with 43.99ms for a local search, 2.30s for a Web search, 3.16s for a Web page browsing, and 12.57ms for a reasoning token. 
In addition, we made the following observations:  
(1) Compared with the parallel search baselines, our method does not significantly increase search and reasoning cost.  
(2) Using the parallel search to integrate knowledge from different sources leads to unnecessary Web search tool calls, which are a lot more expensive and slower than local search tool calls.  
(3) Prompting reasoning models to build deep search agents significantly consumes more reasoning tokens, such as WebThinker and Search-o1. Such token consumption is of limited help for deep search tasks.

\begin{table}[]
    \centering
    \small
    \setlength{\tabcolsep}{0.9mm}{
    \begin{tabular}{lccc|ccc}
    \toprule
    \multirow{2}{*}{\textbf{Method}}  & \multicolumn{3}{c|}{\textbf{Search Success (\%)}} & \multicolumn{3}{c}{\textbf{Reasoning Success (\%)}} \\
    \cmidrule(r){2-4} \cmidrule(l){5-7}
        & Local & Web & Both & Local & Web &  Both \\
    \midrule
        R1-Searcher    & 84.75 & 51.00 & 47.75 & 49.85 & 58.82 & 59.16 \\
        ReCall         & 87.50 & \underline{55.75} & 50.75 & 51.71 & 60.09 & 60.10 \\
        DeepResearcher & 77.25 & 55.00 & \underline{51.00} & \underline{53.72} & 60.45 & \underline{61.27} \\
        Search-o1      & 70.25 & 39.00 & 35.50 & 44.84 & 40.38 & 40.14 \\
        WebThinker     & 71.50 & 52.00 & 48.75 & 47.20 & 57.69 & 56.41 \\
        HierS$_\text{w/o HRL}$ & \underline{89.75} & 23.50 & 22.25 & 51.81 & \underline{62.77} & 61.80 \\
        HierSearch    & \textbf{94.25} & \textbf{81.25} & \textbf{77.75} & \textbf{59.15} & \textbf{63.38} & \textbf{64.63} \\
        
    \bottomrule
    \end{tabular}
    }
    \caption{Multi-Knowledge-Source Utility Analysis on NQ.}
    \label{tab:success-rate}
\end{table}

\begin{table}[]
    \centering
    \small
    \setlength{\tabcolsep}{0.9mm}{
    \begin{tabular}{lcccr|r}
    \toprule
    \textbf{Method}    & \textbf{\# LS} & \textbf{\# WS}  & \textbf{\# WB}  & \textbf{\# Tokens}    & \textbf{Latency (s)} \\
    \midrule
    \textbf{Parallel Search} \\
    R1-Searcher             & 4.26 & 2.13 &  -    & 297.62   & 8.84   \\
    ReCall                  & 4.70 & 2.35 &  -    & 165.15   & 7.70   \\
    DeepResearcher          & 4.42 & 2.21 & 0.01  & 192.57   & 7.72   \\
    Search-o1               & 3.46 & 1.73 & 16.56 & 1,503.71 & 75.43  \\
    WebThinker              & 5.72 & 2.86 & 25.36 & 4,276.77 & 140.83 \\
    \midrule
    \textbf{Selective Search} \\
    CRAG                    & 0.99 & 0.72 &  -    & 1,820.88 & 24.59  \\
    PrefRAG                 & 2.08 & 0.05 &  -    & 1,077.02 & 13.63  \\
    HierS$_\text{w/o HRL}$ & 5.16 & 1.03 & 1.02  & 334.98   & 10.04  \\
    HierSearch             & 3.54 & 1.06 & 2.23  & 408.68   & 14.79  \\
    \bottomrule
    \end{tabular}
    }
    \caption{Efficiency Analysis on MuSiQue.}
    \label{tab:efficiency}
\end{table}

\section{Conclusions}

In this work, we propose a hierarchical agentic paradigm that integrates local and Web searches for enterprise deep search. Our method consists of a low-level local deep search agent and a Web deep search agent that conduct deep search in their corresponding knowledge sources, and a planner agent that coordinates low-level agents and provides the final answer. Furthermore, we devise a knowledge refiner that extracts helpful evidence from low-level agents' trajectories. Extensive experiments demonstrate that our method is effective and efficient across various domains, with better performance in searching and reasoning. This work explores the field of multi-knowledge-source deep search. We anticipate future research questions and research works in this field.


\newpage
\makeatletter
\@ifundefined{isChecklistMainFile}{
  \newif\ifreproStandalone
  \reproStandalonetrue
}{
  \newif\ifreproStandalone
  \reproStandalonefalse
}
\makeatother
\ifreproStandalone
\documentclass[letterpaper]{article} 
\usepackage[submission]{aaai2026}  
\usepackage{times}  
\usepackage{helvet}  
\usepackage{courier}  
\usepackage[hyphens]{url}  
\usepackage{graphicx} 
\urlstyle{rm} 
\def\UrlFont{\rm}  
\usepackage{natbib}  
\usepackage{caption} 
\frenchspacing  
\setlength{\pdfpagewidth}{8.5in} 
\setlength{\pdfpageheight}{11in} 
%
\usepackage{algorithm}
\usepackage{booktabs}
\usepackage{graphicx}
\usepackage{latexsym}
\usepackage{amsmath}
\usepackage{multirow}
\usepackage{array}
\usepackage{comment}
\usepackage{algorithm}
\usepackage{algpseudocode}
\usepackage[tikz]{mdframed}
\usepackage{anyfontsize}
\usepackage{sidecap}
\usepackage{rotating}
\usepackage[table]{xcolor}
\newcolumntype{C}[1]{>{\centering\arraybackslash}p{#1}}
\newcolumntype{L}[1]{>{\raggedright\arraybackslash}p{#1}}

\begin{document}
\fi
\begin{table*}[]
    \centering
    \small
    \begin{tabular}{lllrrrrr}
    \toprule
        \textbf{QA Dataset} & \textbf{Language} & \textbf{Domain} & \textbf{\# Train Set} & \textbf{\# Test Set} & \textbf{\# Text Chunks} & \textbf{\# Entities} & \textbf{\# Triplets} \\
    \midrule
        Musique    & English   & General   & 19.938 & 400  & 135,808 & 968,274   & 1,516,001 \\
        Omnieval   & Chinese   & Financial & 4,769  & 375  & 165,661 & 1,375,359 & 1,858,853 \\
        BioASQ     & English   & Medical   & 35,601 & 340  & 69,292  & 236,697   & 418,836   \\
        NQ         & English   & General   & -      & 400  & 161,022 & 1,121,290 & 1,791,408 \\
        HotPotQA   & English   & General   & -      & 400  & 158,105 & 1,101,900 & 1,774,500 \\
        PubmedQA   & English   & Medical   & -      & 400  & 72,285  & 256,281   & 444,928   \\
    \bottomrule
    \end{tabular}
    \caption{Statistics of all benchmarks. }
    \label{tab:statistics}
\end{table*}

\begin{figure*}
\begin{mdframed}[backgroundcolor=gray!5, roundcorner=3pt, innerleftmargin=10pt, innerrightmargin=10pt, innertopmargin=10pt, innerbottommargin=10pt, nobreak=true]

\small
\sloppy

You are a helpful assistant that can solve the given question step by step with the help of search tools.
Given a question, you need to first think about the reasoning process in the mind and then provide the answer.
During thinking, you can invoke search tools to search for fact information about specific topics if needed.
The reasoning process and answer are enclosed within $<$think$>$ $<$/think$>$ and $<$answer$>$ $<$/answer$>$ tags respectively.
You have access to the following tools:

1. $<$chunk\_search$>$: A dense passage search tool that can be used to search for key passages about specific topics from local corpus.

2. $<$graph\_search$>$: A graph search tool that can be used to search for fact triplets about specific topics.

3. $<$get\_adjacent\_passages$>$: A tool that can be used to get adjacent passages from an entity in the graph.

4. $<$web\_search$>$: A web search tool that can be used to search for fact information about specific topics from the Internet.
\
'/.
5. $<$browse\_url$>$: A tool that can be used to browse a single webpage. You should provide a URL and a question to the tool, and separate them with "|".

The search query for each tool is enclosed within $<$chunk\_search$>$ $<$/chunk\_search$>$, $<$graph\_search$>$ $<$/graph\_search$>$, $<$get\_adjacent\_passages$>$ $<$/get\_adjacent\_passages$>$, $<$web\_search$>$ $<$/web\_search$>$, and $<$browse\_url$>$ $<$/browse\_url$>$ tags respectively.
You should invoke local search tools ($<$chunk\_search$>$, $<$graph\_search$>$, $<$get\_adjacent\_passages$>$) first, and then invoke web search tools ($<$web\_search$>$, $<$browse\_url$>$) if the local search results are not sufficient.

For example, $<$think$>$ This is the reasoning process. $<$/think$>$ $<$graph\_search$>$ search query here $<$/graph\_search$>$ $<$result$>$ search result here $<$/result$>$
$<$think$>$ This is the reasoning process. $<$/think$>$ $<$chunk\_search$>$ search query here $<$/chunk\_search$>$ $<$result$>$ search result here $<$/result$>$
$<$think$>$ This is the reasoning process. $<$/think$>$ $<$get\_adjacent\_passages$>$ search query here $<$/get\_adjacent\_passages$>$ $<$result$>$ search result here $<$/result$>$
$<$think$>$ This is the reasoning process. $<$/think$>$ $<$web\_search$>$ search query here $<$/web\_search$>$ $<$result$>$ search result here $<$/result$>$
$<$think$>$ This is the reasoning process. $<$/think$>$ $<$browse\_url$>$ URL | question here $<$/browse\_url$>$ $<$result$>$ search result here $<$/result$>$
$<$think$>$ This is the reasoning process. $<$/think$>$ $<$answer$>$ The final answer $<$/answer$>$.
Please ensure that all reasoning processes and final answers are enclosed within the correct tags.

\end{mdframed}
\caption{The prompt for the flat deep search agent.}
\label{fig:flat-rl}
\end{figure*}

\begin{figure*}
\begin{mdframed}[backgroundcolor=gray!5, roundcorner=3pt, innerleftmargin=10pt, innerrightmargin=10pt, innertopmargin=10pt, innerbottommargin=10pt, nobreak=true]

\small
\sloppy

You are a helpful assistant that can solve the given question step by step with the help of search tools.
Given a question, you need to first think about the reasoning process in the mind and then provide the answer.
During thinking, you can invoke search tools to search for fact information about specific topics if needed.
The reasoning process and answer are enclosed within $<$think$>$ $<$/think$>$ and $<$answer$>$ $<$/answer$>$ tags respectively.
You have access to the following tools:

1. $<$chunk\_search$>$: A dense passage search tool that can be used to search for key passages about specific topics.

2. $<$graph\_search$>$: A graph search tool that can be used to search for fact triplets about specific topics.

3. $<$get\_adjacent\_passages$>$: A tool that can be used to get adjacent passages from an entity in the graph.

The search query for each tool is enclosed within $<$chunk\_search$>$ $<$/chunk\_search$>$, $<$graph\_search$>$ $<$/graph\_search$>$, and $<$get\_adjacent\_passages$>$ $<$/get\_adjacent\_passages$>$ tags respectively.

For example, $<$think$>$ This is the reasoning process. $<$/think$>$ $<$all\_search\_agent$>$ original question here $<$/all\_search\_agent$>$ $<$result$>$ local search result and web search result here $<$/result$>$ 
$<$think$>$ This is the reasoning process after revising the question. $<$/think$>$ $<$local\_search\_agent$>$ revised question here $<$/local\_search\_agent$>$ $<$result$>$ local search result here $<$/result$>$ 
$<$think$>$ This is the reasoning process after revising the question. $<$/think$>$ $<$web\_search\_agent$>$ revised question here $<$/web\_search\_agent$>$ $<$result$>$ web search result here $<$/result$>$ 
$<$think$>$ This is the reasoning process after revising the question. $<$/think$>$ $<$answer$>$ answer here $<$/answer$>$. 
Make sure to provide a clear and concise final answer based on the most reliable and relevant information.

\end{mdframed}
\caption{The prompt for the local deep search agent.}
\label{fig:local-agent}
\end{figure*}

\begin{figure*}
\begin{mdframed}[backgroundcolor=gray!5, roundcorner=3pt, innerleftmargin=10pt, innerrightmargin=10pt, innertopmargin=10pt, innerbottommargin=10pt, nobreak=true]

\small
\sloppy

You are a helpful assistant that can solve the given question step by step with the help of web search tools. 
Given a question, you need to first think about the reasoning process in the mind and then provide the answer. 
During thinking, you can invoke web search tools to search for fact information about specific topics if needed. 
The reasoning process and answer are enclosed within $<$think$>$ $<$/think$>$ and $<$answer$>$ $<$/answer$>$ tags respectively. 
You have access to the following tools:

1. $<$web\_search$>$: A web search tool that can be used to search for fact information about specific topics from the Internet.

2. $<$browse\_url$>$: A tool that can be used to browse a single webpage. You should provide a URL and a question to the tool, and separate them with "|".

The search query for each tool is enclosed within $<$web\_search$>$ $<$/web\_search$>$ and $<$browse\_url$>$ $<$/browse\_url$>$ tags respectively.

For example, $<$think$>$ This is the reasoning process. $<$/think$>$ $<$web\_search$>$ search query here $<$/web\_search$>$ $<$result$>$ search result here $<$/result$>$.
$<$think$>$ This is the reasoning process. $<$/think$>$ $<$browse\_url$>$ URL | question here $<$/browse\_url$>$ $<$result$>$ search result here $<$/result$>$
$<$think$>$ This is the reasoning process. $<$/think$>$ $<$answer$>$ The final answer $<$/answer$>$.
Please ensure that all reasoning processes and final answers are enclosed within the correct tags.

\end{mdframed}
\caption{The prompt for the Web deep search agent.}
\label{fig:web-agent}
\end{figure*}

\begin{figure*}
\begin{mdframed}[backgroundcolor=gray!5, roundcorner=3pt, innerleftmargin=10pt, innerrightmargin=10pt, innertopmargin=10pt, innerbottommargin=10pt, nobreak=true]

\small
\sloppy

You are a helpful assistant that can solve the given question step by step with the help of search tools. 
You have access to the following search agents:

1. $<$local\_search\_agent$>$: A local search agent that can search for fact information from local corpus and answer questions.

2. $<$web\_search\_agent$>$: A web search agent that can search for fact information from the Internet and answer questions.

3. $<$all\_search\_agent$>$: An agent that invokes both local search agent and web search agent.

Both agents will return evidences, hypotheses, and conclusions. Keep in mind that evidences are reliable but sometimes irrelevant, 
while hypotheses and conclusions are not always correct and may contain hallucinations. 
Given a question, you are first provided with the evidences, hypotheses, and conclusions from both agents. 
The information returned are enclosed within $<$result$>$ $<$/result$>$ tags. 
In case of conflicting information from the two agents, the local search agent's information is considered more reliable. 
You need to carefully scrutinize the evidences, hypotheses, and conclusions from both agents. 
If you find that the information from neither agent is sufficient to answer the question, 
you can revise the question and invoke the search agents again. 
You can invoke the agents using $<$all\_search\_agent$>$ question $<$/all\_search\_agent$>$ for both agents, 
$<$local\_search\_agent$>$ question $<$/local\_search\_agent$>$ for local search agent, 
$<$web\_search\_agent$>$ question $<$/web\_search\_agent$>$ for web search agent, 
and then provide a final answer based on the most reliable and relevant information. 
After you have gathered enough information, you need to provide a final answer. 
The final answer is enclosed within $<$answer$>$ $<$/answer$>$ tags, and enclose your reasoning process within $<$think$>$ $<$/think$>$ tags.

For example, $<$think$>$ This is the reasoning process. $<$/think$>$ $<$all\_search\_agent$>$ original question here $<$/all\_search\_agent$>$ $<$result$>$ local search result and web search result here $<$/result$>$ 
$<$think$>$ This is the reasoning process after revising the question. $<$/think$>$ $<$local\_search\_agent$>$ revised question here $<$/local\_search\_agent$>$ $<$result$>$ local search result here $<$/result$>$ 
$<$think$>$ This is the reasoning process after revising the question. $<$/think$>$ $<$web\_search\_agent$>$ revised question here $<$/web\_search\_agent$>$ $<$result$>$ web search result here $<$/result$>$ 
$<$think$>$ This is the reasoning process after revising the question. $<$/think$>$ $<$answer$>$ answer here $<$/answer$>$. 
Make sure to provide a clear and concise final answer based on the most reliable and relevant information.

\end{mdframed}
\caption{The prompt for the planner agent.}
\label{fig:planner-agent}
\end{figure*}

\begin{figure*}
\begin{mdframed}[backgroundcolor=gray!5, roundcorner=3pt, innerleftmargin=10pt, innerrightmargin=10pt, innertopmargin=10pt, innerbottommargin=10pt, nobreak=true]

\small
\sloppy

\textbf{Question: Who is the sibling of the author of Kapalkundala?}

\textbf{Golden Answer: Sanjib Chandra Chattopadhyay}

\textbf{$<$all\_search\_agent$>$ Who is the sibling of the author of Kapalkundala $<$/all\_search\_agent$>$}

$<$result$>$
Local Chunk Corpus: Kapalkundala Kapalkundala (), also known as Mrinmoyee, is a Bengali romance novel by Indian writer Bankim Chandra Chattopadhyay. Published in 1866, it is a story of a forest-dwelling girl named Kapalkundala ...

Web Page: Kapalkundala - Wikipedia | https://en.wikipedia.org/wiki/Kapalkundala
...
Works of
Bankim Chandra Chattopadhyay
Kapalkundala
(1869) ...
$<$/result$>$

\textbf{$<$think$>$ The author of Kapalkundala is Bankim Chandra Chattopadhyay. His sibling is the answer. $<$/think$>$ $<$local\_search\_agent$>$ Who is the sibling of Bankim Chandra Chattopadhyay? $<$/local\_search\_agent$>$}

$<$result$>$
Local Chunk Corpus: Sanjib Chandra Chattopadhyay Sanjib Chandra was born to an orthodox Brahmin family at Kanthalpara, North 24 Parganas ...
$<$/result$>$

\textbf{$<$think$>$ The author of Kapalkundala is Bankim Chandra Chattopadhyay. His sibling is the answer. $<$/think$>$ $<$web\_search\_agent$>$ Who is the sibling of Bankim Chandra Chattopadhyay? $<$/web\_search\_agent$>$}

$<$result$>$
Web Page: One of his brothers, was also a novelist and is known for his book "Palamau".
Sanjib Chandra Chattopadhyay
Bankim Chandra and his elder brother both went to (then Governmental Zilla School), where he wrote his first poem ...
$<$/result$>$

\textbf{$<$think$>$ The sibling of Bankim Chandra Chattopadhyay is Sanjib Chandra Chattopadhyay. $<$/think$>$ $<$answer$>$ Sanjib Chandra Chattopadhyay. $<$/answer$>$}

\end{mdframed}
\caption{Example trajectory for the planner agent.}
\label{fig:planner-case}
\end{figure*}

\begin{figure*}
\begin{mdframed}[backgroundcolor=gray!5, roundcorner=3pt, innerleftmargin=10pt, innerrightmargin=10pt, innertopmargin=10pt, innerbottommargin=10pt, nobreak=true]

\small
\sloppy

\textbf{Question: Who is the sibling of the author of Kapalkundala?}

\textbf{Golden Answer: Sanjib Chandra Chattopadhyay}

$<$think$>$ First, I need to search for the author of Kapalkundala. Then, I will find the sibling of the author.$<$/think$>$ $<$chunk\_search$>$ Kapalkundala author $<$/chunk\_search$>$ 

$<$result$>$ ...
$<$/result$>$ 

$<$think$>$ The author of Kapalkundala is Bankim Chandra Chattopadhyay. Now, I will search for the sibling of Bankim Chandra Chattopadhyay using the graph search tool.$<$/think$>$ $<$graph\_search$>$ Bankim Chandra Chattopadhyay sibling $<$/graph\_search$>$ 

$<$result$>$ ...
$<$/result$>$ 

$<$think$>$ The graph search tool does not provide information about the sibling of Bankim Chandra Chattopadhyay. I will try using the chunk search tool to find more information about his siblings.$<$/think$>$ $<$chunk\_search$>$ sibling of Bankim Chandra Chattopadhyay $<$/chunk\_search$>$ 

$<$result$>$ ...
$<$/result$>$ 

$<$think$>$ The chunk search tool provides information about the sibling of Bankim Chandra Chattopadhyay. From the search result, it is clear that the sibling of Bankim Chandra Chattopadhyay is Latchmiudayi.$<$/think$>$ $<$answer$>$ Latchmiudayi $<$/answer$>$

\end{mdframed}
\caption{Example trajectory for the local deep search agent.}
\label{fig:local-case}
\end{figure*}

\begin{figure*}
\begin{mdframed}[backgroundcolor=gray!5, roundcorner=3pt, innerleftmargin=10pt, innerrightmargin=10pt, innertopmargin=10pt, innerbottommargin=10pt, nobreak=true]

\small
\sloppy

\textbf{Question: Who is the sibling of the author of Kapalkundala?}

\textbf{Golden Answer: Sanjib Chandra Chattopadhyay}

$<$think$>$ To find the answer to this question, I will need to find the author of the book Kapalkundala and then find information about their siblings. $<$/think$>$
$<$web\_search$>$ the author of the book Kapalkundala and their siblings$<$/web\_search$>$ 

$<$result$>$ ...
$<$/result$>$ 

$<$think$>$ The author of the book Kapalkundala is Bankim Chandra Chattopadhyay. Now I need to find information about his siblings. $<$/think$>$
$<$browse\_url$>$ https://en.wikipedia.org/wiki/Bankim\_Chandra\_Chattopadhyay $|$ siblings of Bankim Chandra Chattopadhyay $<$/browse\_url$>$ 

$<$result$>$ ...
$<$/result$>$ 

$<$think$>$ The author of the book Kapalkundala is not found in the search results. Let me try to find a more specific answer to the question. $<$/think$>$
$<$browse\_url$>$ https://en.wikipedia.org/wiki/Kapalkundala $|$ the author of the book Kapalkundala $<$/browse\_url$>$ 

$<$result$>$ ...
$<$/result$>$ 

$<$think$>$ The author of the book Kapalkundala is Bankim Chandra Chattopadhyay. Now I need to find information about his siblings to answer the question. $<$/think$>$
$<$browse\_url$>$ https://en.wikipedia.org/wiki/Bankim\_Chandra\_Chattopadhyay $|$ siblings of Bankim Chandra Chattopadhyay $<$/browse\_url$>$ 

$<$result$>$ ...
$<$/result$>$ 

$<$think$>$ The author of the book Kapalkundala is Bankim Chandra Chattopadhyay. To find the answer to the question, I need to find specific information about his siblings. Let me look for specific information about Bankim Chandra Chattopadhyay's siblings. $<$/think$>$
$<$browse\_url$>$ https://en.wikipedia.org/wiki/Bankim\_Chandra\_Chattopadhyay $|$ specific information about the siblings of Bankim Chandra Chattopadhyay $<$/browse\_url$>$ 

$<$result$>$ ...
$<$/result$>$ 

$<$think$>$ The author of the book Kapalkundala is Bankim Chandra Chattopadhyay. His siblings are Sanjib Chandra Chattopadhyay and Sarat Chandra Chattopadhyay. Therefore, the answer is Sanjib Chandra Chattopadhyay. $<$/think$>$
$<$answer$>$ Sanjib Chandra Chattopadhyay $<$/answer$>$

\end{mdframed}
\caption{Example trajectory for the Web deep search agent.}
\label{fig:web-case}
\end{figure*}

\begin{figure*}
\begin{mdframed}[backgroundcolor=gray!5, roundcorner=3pt, innerleftmargin=10pt, innerrightmargin=10pt, innertopmargin=10pt, innerbottommargin=10pt, nobreak=true]

\small
\sloppy

\textbf{Question: Who is the sibling of the author of Kapalkundala?}

\textbf{Golden Answer: Sanjib Chandra Chattopadhyay}

$<$think$>$ To answer this question, I need to find the author of Kapalkundala and their sibling. I will use the graph search and chunk search to find the author of Kapalkundala. Then, I will use the chunk search and get\_adjacent\_passages to find the sibling of the author. $<$/think$>$ $<$graph\_search$>$ author of Kapalkundala $<$/graph\_search$>$ 

$<$result$>$ ...
$<$/result$>$ 

$<$think$>$ The author of Kapalkundala is Bankim Chandra Chattopadhyay. Now, I will use the chunk search to find the sibling of Bankim Chandra Chattopadhyay. $<$/think$>$ $<$chunk\_search$>$ sibling of Bankim Chandra Chattopadhyay $<$/chunk\_search$>$ 

$<$result$>$ ...
$<$/result$>$ 

$<$think$>$ The chunk search did not provide any information about the sibling of Bankim Chandra Chattopadhyay. I will use the get\_adjacent\_passages to find the sibling of Bankim Chandra Chattopadhyay. $<$/think$>$ $<$get\_adjacent\_passages$>$ Bankim Chandra Chattopadhyay $<$/get\_adjacent\_passages$>$ $<$result$>$

$<$/result$>$ $<$think$>$ The get\_adjacent\_passages did not provide any information about the sibling of Bankim Chandra Chattopadhyay. I will use the chunk search again to find the sibling of Bankim Chandra Chattopadhyay. $<$/think$>$ $<$chunk\_search$>$ sibling of Bankim Chandra Chattopadhyay $<$/chunk\_search$>$ 

$<$result$>$ ...
$<$/result$>$ 

$<$think$>$ The chunk search did not provide any information about the sibling of Bankim Chandra Chattopadhyay. I will use the web search to find the sibling of Bankim Chandra Chattopadhyay. $<$/think$>$ $<$web\_search$>$ sibling of Bankim Chandra Chattopadhyay $<$/web\_search$>$ 

$<$result$>$ ...
$<$/result$>$ 

$<$think$>$ The web search did not provide any information about the sibling of Bankim Chandra Chattopadhyay. I will use the browse\_url to find the sibling of Bankim Chandra Chattopadhyay. $<$/think$>$ $<$browse\_url$>$ https://www.britannica.com $|$ Bankim-Chandra-Chattopadhyay - Encyclopædia Britannica $<$/browse\_url$>$ 

$<$result$>$ ...
$<$/result$>$ 

$<$think$>$ The browse\_url did not provide any information about the sibling of Bankim Chandra Chattopadhyay. I will use the graph search again to find the sibling of Bankim Chandra Chattopadhyay. $<$/think$>$ $<$graph\_search$>$ sibling of Bankim Chandra Chattopadhyay $<$/graph\_search$>$ 

$<$result$>$ ...
$<$/result$>$ 

$<$think$>$ The graph search did not provide any information about the sibling of Bankim Chandra Chattopadhyay. I will use the chunk search again to find the sibling of Bankim Chandra Chattopadhyay. $<$/think$>$ $<$chunk\_search$>$ sibling of Bankim Chandra Chattopadhyay $<$/chunk\_search$>$ 

$<$result$>$ ...
$<$/result$>$ 

$<$think$>$ The chunk search provided information about the sibling of Bankim Chandra Chattopadhyay. The sibling of Bankim Chandra Chattopadhyay is Kanaka Durga. $<$/think$>$ $<$get\_adjacent\_passages$>$ Kanaka Durga $<$/get\_adjacent\_passages$>$ $<$result$>$

$<$/result$>$ $<$think$>$ The get\_adjacent\_passages confirmed that the sibling of Bankim Chandra Chattopadhyay is Kanaka Durga. $<$/think$>$ $<$answer$>$ Kanaka Durga $<$/answer$>$

\end{mdframed}
\caption{Example trajectory for the flat-RL-trained deep search agent.}
\label{fig:flat-case}
\end{figure*}

\section{appendix}

\subsection{Prompts}

We provide the prompts used in HierSearch's training and inference processes. These include prompts for the flat-RL-trained deep search agent (Figure~\ref{fig:flat-rl}), local deep search agent (Figure~\ref{fig:local-agent}), Web deep search agent (Figure~\ref{fig:web-agent}), and planner agent (Figure~\ref{fig:planner-agent}). Some agents' prompts refer to the prompt design of ReCall.\footnote{\url{https://github.com/Agent-RL/ReCall}} 

\subsection{Benchmarks}

In Table~\ref{tab:statistics}, we provide the number of training samples (\# Train Set), test samples (\# Test Set), text chunks in the local retrieval library (\# Text Chunks), as well as the number of entities (\# Entities) and triples (\# Triplets) in the local knowledge graph for the 6 datasets we used.

For local search, we prepare local knowledge bases separately for general, medical, and financial domains. For the general domain, we sample passages from the Wikipedia dump, and for the medical domain, we sample passages from the PubMed dump. 
\footnote{Wikipedia dump: \url{https://huggingface.co/datasets/RUC-NLPIR/FlashRAG_datasets/tree/main/retrieval-corpus}. PubMed dump: \url{https://huggingface.co/datasets/MedRAG/pubmed}.}

\subsection{Implementation Details}

We collect 19,938 training samples from Musique, 4,769 samples from OmniEval, and 35,601 samples from BioASQ for agent training. We train the local deep search agent, Web deep search agent, and the planner agent for 300 steps with a batch size of 64. We use the GRPO algorithm implemented in the VERL framework
\footnote{VERL: \url{https://github.com/volcengine/verl}.}. We set the learning rate to 1e-6, batch size to 64, and train on an 8-GPU H20 machine. We choose \texttt{Qwen2.5-Instruct} as the backbone model for both search agents and the planner agent, which is the backbone model for RL-related baselines~(R1-Searcher, ReCall, and DeepResearcher).

\subsection{Case Study}

We further analyze deep search performances of our method and the agent trained by flat RL in multi-knowledge-source environments through case studies. Figure~\ref{fig:planner-case}, Figure~\ref{fig:local-case}, and Figure~\ref{fig:web-case} show the thinking trajectories of agents used in HierSearch. Figure~\ref{fig:flat-case} shows the trajectory of the flat-RL-trained agent. The question-answer pair we selected is from MuSiQue. Local search tools can only access part of the knowledge, while web search tools can supplement the knowledge. The flat-RL-trained agent is not proficient in using web search tools, leading to incorrect answers. In our method, the local agent found some helpful evidence, and the web agent successfully supplemented the missing knowledge. Finally, the planner reaches the correct answer.

\ifreproStandalone
\end{document}
\fi

\end{document}